# A framework for adaptive real-time applications: the declarative real-time OSGi component model

*Ning Gui*   *Vincenzo De Florio*   Hong Sun   *Chris Blondia*

University of Antwerp
Dept. of Math and Comp. Science, PATS group,
Middelheimlaan 1, 2020 Antwerp, Belgium,
and IBBT,Ghent-Ledeberg, Belgium
{ning.gui, vincenzo.deflorio,hong.sun, chris.blondia}@ua.ac.be

## ABSTRACT
Nowadays, more and more applications require OSGi to have some form of real-time support, which is currently very limited. The resulting closed-system solutions lack of a standard management scheme which forbids standard, system-wide policies for real-time system's deployment, adaptation, and reconfiguration. In order to tackle this problem, this paper proposes a declarative real-time component model. In this model, the distinguishing real-time contract of each component is declaratively described, and a general component real-time management interface is designed. They are used to maintain an accurate view of existing real-time components' promised contracts. A real-time component runtime service is designed to control the whole lifecycle of the components. By using global information and general control interface, it can adjust the system continue to operate without impairing the deployed components' real-time contracts in the face of run-time changes. This system allows itself to be easily extended with other constraint resolving policies to fit different context. The prototype has been tested into a simulated control system. The result shows this framework can provide good real time performance while still provides real-time component dynamicity support as well. To the best of our knowledge, this is the first comprehensive solution providing explicit real-time support from design to execution in OSGi framework.

## 1. INTRODUCTION
The OSGi [1] service platform is the most widely adopted technology to build a control system for the networked home. Many features contribute to its success. To name a few, it supports various well-known protocols, simplifying communication among home devices, such as UPNP. In addition, it defines a service oriented model for applications to dynamically discover and use services provided by others. It allows flexible remote management of these applications and the services they provide. But, most importantly for our discussion herein, it designs a flexible continuous deployment platform which supports the system to install, update, and uninstall the bundles without restart the whole system. An OSGi solution supports decentralization, diversification, and ubiquity not only in smart homes but in pervasive environments and even business context generally.

However, the OSGi specification lacks system-wide support for real time applications. This absence greatly limits its application in many contexts where specific real-time (soft or hard) requirements exist. An example is given by the Set-Top Boxes needed to decode/encode media data, which has typical soft real-time characteristics. This also restricts OSGi's application in industrial control environments, where the hard real-time merits need to be guaranteed. Some approaches such as Real time Java specification [2] to some extent mitigate such problems by using special real-time thread classes to perform real-time tasks. However, in complex real-time systems, these tasks normally are, to a certain extent, inter-dependent with one another, as one real-time task may have certain impacts on other tasks' real-time performance while competing for the CPU, memory or other system resources. Ad-hoc solution results in lack of accurate global view of the existing real-time context. This absence may lead to breaking the real time contracts which the system designer has supported to satisfy [3]. The continuous deployment support in OSGi platform actually exacerbates this problem, for the bundles (component) can be installed, started, stopped and uninstalled at run-time. That means the traditional waterflow approach: design, verify, map, and deploy, could not work effectively because the system configuration will evolve during the whole system lifecycle. In other words, the static system configuration assumption is not valid in typical OSGi system.

The motivation for this research is to explicitly support dynamic availability of real time components in OSGi and simplify the real time component constraints resolving. We designed a Declarative Real-time Component Executive (DRCR) service to solve the constraints between real time components at run-time. This service mitigates the burden for individual component implementations on monitoring changes and constraint resolving. Changes in system configuration and real time block composition are under control of DRCR runtime rather than individual applications. In order to cope with different system contexts, we also designed a resolving service to provide customized real-time admission and adaptation service, which can be plugged into the DRCR runtime by using OSGi service model.

The rest of this paper is structured as follows: in Section 2 we summarize the challenges related to using OSGi in a real-time context. In that same section we also describe the principles of the declarative real time component model and analyses the lifecycle management of the declarative real-time components. Section 3 then introduces our prototype implementation. Section 4 describes the scenario execution and report the performance of our implementation, followed in Section 5 by a comparison with related works. Section 6 presents future work and the conclusions..

## 2. OSGi & Real Time component model

The OSGi specification defines a service platform that includes a minimal component model, a small components managing framework, and a service registry. However, even with its extensible and flexible component model, OSGi still faces many challenges in dealing with real-time requirements. In this section, we will analyze the challenges that may emerge when bringing explicit real-time support to existing OSGi frameworks and introduce the Declarative Real-Time Component Model.

### 2.1. Challenges – OSGi in Real-Time Domain

Although OSGi and its extension – Declarative Service [5] – provide a very flexible component model, they still face many challenges in dealing with real-time requirements. Here, we describe those challenges and discuss the shortcomings in OSGi in dealing with these real-time requirements.

**Dynamic availability** of real-time components results for any number of reasons, such as context awareness computing or continuous deployment activities. Since these changes of availabilities occur at run time, they have repercussions on existing real time component instances and the compositions in which they reside. Addressing dynamic availability is a complex task that requires component instances to monitor for status changes, and asks applications to listen for component departure or arrival. This adds extra burden to the real-time component developer's already error-prone and arduous work.

**Real-time component composition**: component composition is one of the key issues of the component based system. Its correctness in construction is based on component's information, which is needed to check for composability with other components. OSGi framework provides a generic and standardized solution for Java modularization. However, composition of modules is still largely based on import and export of java packages resolved by the LDAP filter. This approach tightly couples with the Java language and could not specify domain-specific component relationships. Although, from OSGi 4.0, the declarative service was introduced to support the dynamic composition of service oriented components, it still tightly coupled with Java language. Furthermore, the policy for service matching is predefined and static, whereas the requirements of real-time applications are normally very complex and application specific. Certain extensible mechanism is needed to support such complexity.

**Lack of global view of real time conte**xt limits the usage of the real time application in OSGi. In real time systems, a task' execution characteristics are often data-dependent, configuration-dependent and system-dependent. The sets of execution tasks and available processor resources may not be known until run time, and may change in the run-time. Thus, ensuring component real-time contracts is not simply a matter of compile-time guarantees that the provided and required functions are accessible regardless of the software module's location and that those formal and actual parameters match. It also demands that a noninterference mechanism guarantees that all deployed essential components' real-time properties are preserved during system construction and evolution [3]. However, in the OSGi framework, there is no such specific mechanism for global resource management, such as CPU usage management, memory utilization control. Due to the fact that these resources are globally shared, the resource budget should be "enforced" by a central scheme rather than by each single bundle (component).

Many existing real-time component specifications only address issues related to application-specific real-time guarantees and dynamicity, which thus underestimates the complexity of current and future real-time applications. These applications' component interactions depict a variety of timing dependencies due to scheduling, communication, synchronization, arbitration, blocking, and buffering. So the challenges of implementing real-time behavior include not only decoupling and modularizing their behavior, but also the ability to maintain a global view of existing real-time components' promised contracts and requirements to ensure that the system continues to operate in the face of events that occur at run time. If a system fails to address this problem, the composition will eventually lead to possibly transient timing problems, including missed deadline, task and message skipping, or overwriting and buffer overflows.

### 2.2. Declarative Real-Time Component Model

In order to cope with the challenges as stated above, in this paper we introduce an explicit real-time component support for existing OSGi framework – the Declarative Real-time Component (DRCom) model. A DRCom is a normal Java class and corresponding real-time code contained within a bundle. The distinguishing real-time aspect of DRCom is declared in an XML document which describes the real-time related information. This information consists of basic real-time component information such as the name and type of task, priority, and frequency. It also contains outgoing ports descriptions that are implemented by the component, as well as incoming ports which specify the component's functional dependence. This description can be easily extended to describe more complex configurations. When the component is deployed into the system, the DRCR service will automatically parse its real-time component configuration and store these data into its internal registry.

The DRCR execution environment manages the real-time component's dynamicity according to the current system context. Component configurations are activated and deactivated under the full control of DRCR which holds the global view of all real-time components installed in the current system. Decisions were made based on the information about current system context, and possible with the help of a user-customized resolving service, if exist. By taking into account each component's real-time contract and the current system configuration, we provide the basis to support the component's dynamicity without impairing the deployed component's real-time contracts. Component's real-time contracts are now guaranteed by the execution environments rather than by each component itself. One key feature of our framework is that DRCR controls the real time component's lifecycle. It is only through this approach that the DRCR can keep a complete and accurate global view of current system context. Doing otherwise, i.e. allowing each component to be created or destroyed by its own proprietary interfaces/methods, the system would lose track of the deployed components' state and system resource utilization status. The resulting incomplete and inaccurate information will definitely hinder the system ability to cope with such dynamicity.

The dynamic behavior of the declarative real-time component is as shown in Figure 2. As the Declarative Real-time Component model is based on the OSGi bundle, so its lifecycle is a sub life-cycle of OSGi bundle. After the OSGi bundle is enabled by the OSGi framework and contains valid declarative real-time component description, the DRCR will take control of its lifecycle and manage its real-time related requirements. In this model, parts of lifecycle control are driven by external events such as component deployment and destruction (which

```xml
<? xml version="1.0" encoding="UTF-8"?>
<drt:component name="camera" desc="this is a smart camera controller" type="periodic" enabled="true" cpuusage="0.1">
    <implementation bincode="ua.pats.demo.smartcamera.RTComponent"/>
    <periodictask frequence="100" runoncup="0" priority="2"/>
    <outport name="images" interface="RTAI.SHM" type="Byte" size="400" />
    <inport name="xysize" interface="RTAI.SHM" type="Integer" size="400"/>
    <property name="prox00" type="Integer" value="6" />
    …
</drt:component>
```

Figure 2: Sample configuration for real-time task

still need to go through DRCR). Some state changes are automatically managed by DRCR, such as *Unsatisfied* and *Active*. During execution, the DRCR receives notifications from the OSGi framework for component state changes. These notifications can trigger re-configuration activities, which depend on the system context and the real-time properties associated to the component that is being managed. The behavior of declarative real-time component is shown as Figure 1.

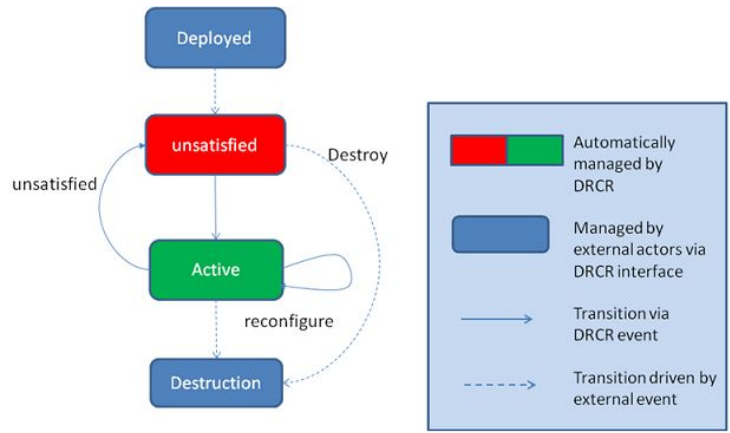

Figure 1: Declarative Real-time Component Lifecycle

## 2.3. Real-Time Component Description

In DRCom implementation, a real time component description represents a component's real time requirements and functional contracts. The DRCom description contains the real time task's name which is used to create tasks in the real time domain, as well as the task's type and descriptions. For periodic task, it can also specify priority, CPU allocation and task's performance frequency. The description can also contain a set of component specific properties which can be used to configure and identify the component instance from those of other providers. In order to communicate with other real-time components, the component should also specify a set of ports that represent the communication interfaces. Figure 3 shows a fragment of meta-data file which describes a smart camera that can return regions of interests (subsets from a frame image data) on demand. Such model is being used in the framework of IST project ARFLEX [16].

The **component element** has the following attributes:

• *name*: The name of a component must be globally unique because it is used as a task reference in several places. In current implementation, the ports are characterized by a six character name because the underlying real time OS use the six character name to refer to the real time tasks.

• *enabled*: Controls whether the component is *enabled* when the bundle is started. The default value is true. If enabled is set to false, the component is disabled until the method enableRTComponent is called.

- *type*: specifies the real time component task type. Its value can be periodic and aperiodic. If the task's type is set, the corresponding elements should also be included in the component description. In Figure 2, the task's period is set to 10 millisecond and is to run on CPU 0 with priority 2.
- *cpuusage*: using this attribute, the component can specify how much CPU it will claim to guarantee its real-time characteristics.

The **implementation element** is required and defines the name of the component implementation class. It has therefore only a single attribute: bincode. It's a Java fully qualified name of the implementation class. The component instances will be created by DRCR by referring to this attribute.

The **inport** and **outport elements** define the communication methods from which the inter-components data are shared. The component may have 0 or more inports or outports. The inport and outport have nearly identical attributes except the data transmission direction.

**inport(outport):** specifies what the required input (output) for the real-time component is. It consists of a *name* attribute, which is also used as the communication reference. In addition to name, the input port should have

- *interface*- the communication interface. In current prototype, only the RTAI.SHM (shared memory) and RTAI.Mailbox are supported. • *type*- the transported data type,which can be integer or byte.. • *size*- the number of data: it is the multiple size of the data type's size.

Together with the name attribute, these attributes are used to determine the port compatibility between the provided and required interfaces.

A component description can define a number of **properties**. The properties can be used to configure the real-time component while it can also be used to identify the component from other providers. The property element has the attributes name, value and type, in typical Java configuration format.

## 2.4. DRCom management Interfaces

In order to achieve a coherent way to control basic component behavior, each compatible real-time component is required to implement the real-time component management interface. This interface will be registered as management service by DRCR together with the component's properties in the service registry of OSGi. It can be discovered dynamically and allow other OSGi modules to participate in the dynamic reconfiguration activities to fine tune the system performance. While still provide the basic component management service, we try to keep the interface as simple as possible. The current management interface defines the methods to suspend, resume, get/set properties and get status of a real-time task.

By search in the service registry, general component's user can locate the individual component and use this service to control the component's behavior or get certain information from it, General or application specific adaptation managers can monitor the tasks status and adjust the parameter or even change the application structure according to current available resources and system requirements.

Please note: In order to keep a accurate global view, although the component implement the *init* and *uninit* methods, they are not exposed in the component's interface.

## 3. Prototype Implementation

In control systems, each component can be mathematically modeled using a transfer function, which computes an output response for any given input [6]. However, with the growing complexity of real-time systems, the real-time component is characterized by more and more non real-time related requirements. These aspects include(but not least) the deployment process of components and the configuration and reconfiguration phase of components.

## 3.1. Hybrid Real-time Component approach

In complex real-time systems, real-time capabilities are often not necessary for the whole large applications, but only for small parts of them. In this framework, we propose the Hybrid Realtime component (HRC) implementation model which consists of two main parts – one part runs directly in the real-time OS layer and the other part runs in the OSGi middleware environment. The result is a split architecture where we have a large non-real-time container, which is based on OSGi. For the real-time OS support, an operating system that allows its modification and also offers hard real-time characteristics is needed. Therefore, open source real-time Linux kernel extension- RTAI- was chosen. The figure 3 shows the architecture of this implementation.

The real-time part of each HRC is an independent concurrent process, whose functionality is defined by the methods of a standard object. The standard object implements a set of functions that enable it to receive data and commands from the non real time parts while response status inquiry from its non real-time partner. Each real-time task has certain general properties such as name, description, task type etc, and task specific properties, which are used to (re)configure the real-time component for use with specific hardware or applications. Communication with other real-time modules is restricted to its input and output ports. The real-time task can also connect to sensors or actuators, via the digital I/O module. The details of accessing the hardware are encapsulated within the real-time task.

In the non real-time part, we implemented a configuration specific interface containing methods for getting/setting component parameters or getting the component state. The non real-time part is implemented in line with OSGi specification. Its management interface is registered in the system registry and thus can be managed by external modules. We reused the OSGi framework service to realize the modules deployment, version control, etc.

A key concept in our approach is that we want to separate the actual adaptation logic as much as possible from the real time business logic constituting the components' code. The large and complex adaptation and management parts run in a classic non-realtime environment and only small, predictable parts in a real-time environment. This component model is rather concise and easy to implement compared to the pure real-time component model.

We also designed the formalized intra communication interfaces working as the bridge between real-time code and its non real-time count part as well as the inter component communication interfaces [9]. In current stage, the implementation is just meant to show the feasibility to implement the declarative real-time component model and provide the related performance results for quantitative analysis. It is still in tentative stage and only limited component communication methods are provided. Here, although our implementation use the Hybrid real-time implementation approach, the DRCom model don't have any specific implementation method requirements as long as it support the real-time management interface, that means it can easily adapt other possible implementation.

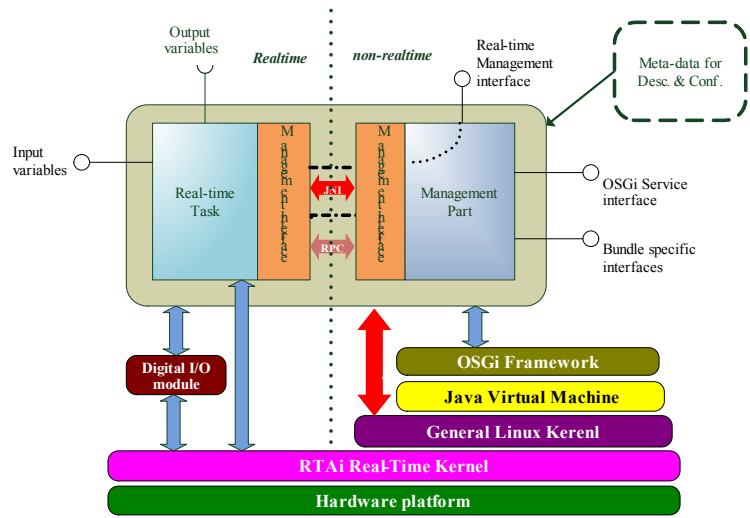

Figure 3: split container architecture running the Hybrid real-time component.

### 3.2. Intra-component communication

Due to the hybrid structure of our HRC, the intra communication interface needs to be standardized to simplify the development of the component. We designed one interface that needs to be implemented by the OSGi parts and its real-time code. This interface works orthogonally to the functional part of the real-time task. The minimal communication interface has to incorporate the basic methods defined in the component management interface described in section 2.4.

Through the intra-component's communication, the non-realtime part can monitor the status of the real-time component code. In the meantime, any change in the OSGi lifecycle corresponds to an appropriate reaction in the real-time code's lifecycle, maintaining the coherence among both frames.

In the prototype, in order to keep the real-time task's real-time behavior, real-time code should not wait for the command sent by the non real-time countpart. Asynchronized communication mode was chosen as the basic communication methods between real-time and non-real-time part. Otherwise, the real-time task's performance may be breached. The prototype currently uses the RTAI's Inter process call (IPC) as the command transmission method. When the task finishes its main functional routine, it tries to read command message sent asynchronously through the management interface. Implementation is described in details in [9].

### 3.3. Inter-realtime component communication

Pure OSGi register based service reference location may not handle the real time invocation timely and efficiently. While there is some proposals such as [15] that want to provide real-time registry search support and in OSGi, their plan still could not solve the problem of interface based component inter-action which do not have any real-time guarantee. In order to solve this problem, they try to use at ARINC 653 OS Partitioning to prevent Bundles from interfering with one another. It actually needs to revamp the whole OSGi system implementation.

In our implementation, the DRCR just provides the component's real-time resolving and lifecycle management support. Inter- realtime communication is directly mapped to the real-time OS container – the RTAI based real-time runtime. RTAI provides a set of efficient real-time inter-process communication and resource management support. In order to keep tasks' real time performance, the non real-time OSGi implementation will not directly interfere with the inter task communication. This approach will keep the existing OSGi implementation largely intact while still providing very good real-time communication support. The simulation results also show the efficiency of this approach.

### 4. Execution Scenario & Performance Evaluations

As we know that one of the basic features in our designed systems is the dynamicity of declarative real-time components. Here we use a simple scenario to show the adaptability. And we measure the performance of this system in terms of the latency of scheduling the periodic real time task with/without the declarative real-time component support.

## 4.1. Test Environment

In our experiments, we use one HP nc6400 laptop which has 1.66 GHz CPU duo core T5500, 1.5GB RAM and 160 GB HDD. The software system is run on the Fedora 7 kernel 2.6.20 with RTAI Patch version 3.5. The Declarative real time service implementation is based on Equinox (Version 3.2.1), which is a popular, free, open source OSGi Platform developed by Eclipse organization. The JVM we adopted is JAVA 1.6.0.2 SDK. The scheduler used in the test is round-robin algorithm. In the implementation, we use the RTAI LXRT module– which allows the use of the RTAI system calls from within standard user space.

## 4.2. Component configuration

The application is converted from the RTAI's system performance test suit for best demonstrate the system's adaptability and hybrid implementation performance of the declarative real-time component model. The application consists of two real-time components which are delivered as individual bundles. One of two components will do some simulated computing job at rate of 1000Hz, the *calculation* task and the other task- *display* task, will display the scheduling latency at rate 4 by reading the shared memory. Due to functional constraint, the display task will rely on the calculation task's outport, so it could not start if no active calculation task exists. Here, in order to the test concise, we supports all the other resource constraint will be satisfied.

## 4.3. Evaluation of Dynamicity

Here, we use different scenarios to demonstrate how the system supports component dynamicity.

In this scenario, the Component Display needs component Calcuation's output to satisfy its functional constraints. As an instance of component Calcuation already exists, the DRCR will successfully solve Display's functional constrains(its functional constraint is satisfied). Then, the internal resolving service and the external customized service will be consulted for real-time non –functional constraints resolving process. When both services return positive results, (in the simulation, both results is true) the DRCR will create and activate the component Display's instance. While if component Calcuation is stopped, the DRCR gets notified about this event and consults its internal resolving service and the external customized service again to check for possible unsatisfied component instances. When get the results, the DRCR will find component Display's instance is unsatisfied and should be disabled. Due to page limits, the figures of the whole process could not be list here.

## 4.4. Simulation results

Performance was evaluated in two implementations, the Pure RTAI user model and the DRTCom model in verifying the architectural latency. In both models, the real-time task is coded as the RTAI user model tasks. The hardware Timer is set in periodic mode. The reasons for such a test are that there will be always a drift between time baseline and the one the task are really scheduled. The difference and jitters in certain extent reflect the real-time system's performance.

We test the performance in two different system load environments (light & stress mode) and the system runs the two real-time tasks in two implementations: pure RTAI and our hybrid implementation. You can see the latency result in the declarative component mode actually has no much difference with the application in pure RTAI environments. It is quite encouraging because it means this implantation can provide the latency guarantee in 30us which is quite nice for much real-time application.

We also test the performance in stress model. In this mode, we use the following three commands accompany with our OSGI platform. The CPU usage is close to 100%.

**Table 1 Latency Test (light & Stress) mode**

|                  | AVERAGE   | AVEDEV  | MIN     | MAX    |
| ---------------- | --------- | ------- | ------- | ------ |
| HRC (light)      | -1334.9   | 3760.03 | -24125  | 21489  |
| Pure RTAI(light) | -633.8    | 3682.82 | -25436  | 23798  |
| HRC (stress)     | -21083.74 | 338.89  | -23314  | -17956 |
| Pure RTAI(stress)| -21184.52 | 385.41  | -25233  | -18834 |

From table 1, we can also see the hybrid implementation has the similar performance whenever the system is in stress mode or light work mode. It's because the real-time functions are implemented as RTAI task which has higher priority than normal Linux process due to the fact that RTAI actually use dual-kernel approach. With the help of this approach, it solves one of the biggest challenges in this context is to prevent Java's garbage collector from interfering with real-time task scheduling

## 5. Related work

Our platform follows the approach of *descriptive* real-time support for OSGi framework; that is, the real-time guarantee is not directly implemented within components' application code but provided implicitly by the container runtime environment according to additional component descriptors.

CIAO [9] builds a QoS-enabled CORBA component model implementation on top of TAO [10]. The project adhered to existing OMG specifications such as RT/CORBA and CORBA Component Model (CCM) and the extension of those. The considerable overhead of implementing or extending a fully compliant CCM infrastructure would have been counterproductive to our system goal.

In the real-time component designs, Stewart et al. [11] designed a reconfigurable port-based object framework which has similar real-time communication scheme as ours. But their design used a pure real-time design philosophy which makes this system hard to develop.

In order to deal with component dynamicity, Cervantes,H. and Hall,R.S [12] propose a service-oriented component based framework for constructing adaptive component-based applications. The key part of the framework is the Service Binder which automatically controls the relationship between components. Our approach mimics theirs in dealing with real-time component's dynamicity. However, we add the Customer Components Resolving Service which enables the application specific adaptation. It limits its application in normal OSGi applications and could not be applied to the real-time domain.

Robert W. Brennana etc, [13] propose a reconfigurable concurrent function block model based on real time Java. Their component model is based on IEC 61499 standard. Their approach tries to tackle the dynamic configuration problems but their approach lacks of the declarative real-time component descriptions and the relationship is manually managed by a so-called Function block manager interface. It lacks the automatic component lifecycle management and its prototype actually requires people to monitor and control the process.

Hartig and Zschaler designed and implemented enforceable component-based real-time contracts [14] – the extension of component-based software engineering technology that comprehensively supports adaptive real-time systems from specification all the way to the running system. Their work is the most similar approach to our framework. Their implementation is based on JBoss and a small real-time container (CONQOS RT). However, although they propose the concept of adaptation manager for parameter adjustment and profile change, there is no formal design for how to deal with the dynamicity of component's availability which is crucial for downtime-free systems.

## 6. Conclusion and future work

This paper described the experience of building real-time component model in OSGi that supports run-time adaptation in response to the dynamic evolution and continuous deployments of modern complex real-time systems. In this framework, the real-time contract is specified in the component's meta-data. The component instance was managed by the system for the dependence resolving and real-time contract. Global view of current system configuration is managed by the DRCR service. Such system reasons about the changes in the system configuration and perform certain actions while still guaranteeing the activated real-time components' real-time contracts. It sheds the burden for each real-time component to monitor the system status and maintain the reference to the other dynamic available real-time components.

Our capability to give real-time guarantees is based in the RTAI, which provides schedulers for RTAI tasks which have higher priorities than normal Linux tasks via its dual kernel approach. The gap between the component-based specification using declarative component specification and the task-based world of RTAI is bridged by a hybrid component implementation approach and the DRCR. This DRCR decides how to use the available resources for the components to be executed. The implementation we have presented is split into two parts: (a) a large part based on OSGi, which performs complex component management, and (b) a small real-time-capable part running directly on the real-time operating system and executing components with real-time demands. Hence, the dynamic adaptation of component-based applications to changes could be performed at runtime and the component developer does not need to bother about this management code.

Although we were able to show an integrated approach to providing real-time properties to component-based applications, areas remain open for further research:

At the moment, the semantics of real-time component specifications define only the basic parts of the real-time requirements. To allow for Declarative component specifications to be integrated into development processes even more complex than the current ones, more powerful component description language is needed.. There is limited communication support between real-time tasks. In order to support such requirements, the full-fledged semantics of DRCom must be careful envisaged and defined. We are working to integrate certain Architecture Description Language into our DRCom. In the meantime, more complex real-time component implementation should be designed to cope with the port based components' limitations.